\documentclass[twocolumn,aps,prl,showpacs]{revtex4}
\usepackage{times}
\usepackage{graphics}
\usepackage{graphicx}

\usepackage{dcolumn}

\usepackage{amsmath}
\usepackage{epsfig}
\usepackage{epstopdf}

\usepackage{ulem}
\usepackage{color}

\begin{document}
\bibliographystyle{apsrev}

\title{Millikelvin Spatial Thermometry of Trapped Ions}
\author{B.G. Norton}
\email{benjamin.g.norton@gmail.com}
\author{E.W. Streed}
\author{M.J.Petrasiunas}
\author{A. Jechow}
\author{D. Kielpinski}

\affiliation{Centre for Quantum Dynamics, Griffith University, Nathan QLD 4111, Australia}
\date{\today}

\begin{abstract}
We demonstrate millikelvin thermometry of laser cooled trapped ions with high-resolution imaging. This equilibrium approach is independent of the cooling dynamics and has lower systematic error than Doppler thermometry, with $\pm 5$ mK accuracy and $\pm 1$ mK precision. We used it to observe highly anisotropic dynamics of a single ion, finding temperatures of $<60$ mK and $>15$ K simultaneously along different directions. This thermometry technique can offer new insights into quantum systems sympathetically cooled by ions, including atoms, molecules, nanomechanical oscillators, and electric circuits.
\end{abstract}


\maketitle
\newpage

Laser-cooled trapped ions are a nearly ideal system for investigations of quantum physics. The internal ion states are strongly decoupled from the surrounding environment and can exhibit coherence times of many seconds. In ultrahigh vacuum the trapped ion motion is strongly coupled through the Coulomb force but otherwise exhibits good immunity to external perturbations. Precision manipulation of ions at the quantum level is readily achieved through the use of laser and electromagnetic fields. These properties have made laser-cooled trapped ions a preferred platform for implementing experiments in quantum information processing (QIP) \cite{Ripoll-Cirac-2005,Amini-Wineland-2010,Kielpinski-2003,Kim-2010}, and precision metrology\cite{Chou-Wineland-2010,Biercuk-Bollinger-2010}. The strong Coulomb coupling makes laser-cooled trapped ions attractive for sympathetic cooling at millikelvin temperatures in investigations of fundamental physics \cite{Schuster-11}, the dynamics of complex molecular \cite{Ostendorf-06} and biomolecular \cite{Offenberg-08} ions, nano-mechanical oscillators\cite{Tian-2004,Hensinger-2005}, resonant electric circuits\cite{Rabl-2010}, and Bose-Einstein condensates\cite{Zipkes-10}.

Thermometry is a key diagnostic in all of these experiments. Effective Doppler cooling is a precondition for QIP and precision measurement applications.  The development of new ion trapping architectures for QIP requires millikelvin thermometry to quantify high anomalous heating rates\cite{Labaziewicz-08,Daniilidis-11}. Motional decoherence is a limiting factor in recent experiments including the long range mechanical coupling of ions\cite{Harlander-2011, Brown-2011} and the realization of a phonon laser \cite{Vahala-09}. In sympathetic cooling, thermometry of the cooling ions is the primary diagnostic for the temperature of the cooled species. In the presence of a controlled heating process, thermometry of the cooling ions can be used to measure the cooling power of the sympathetic cooling process. Sympathetic cooling is a necessary precondition for mixed species experiments such as the Al$^+$/Be$^+$ ion clock \cite{Chou-Wineland-2010} or molecular ion experiments\cite{Schuster-11,Ostendorf-06,Offenberg-08}. These applications of thermometry are especially useful for proposed hybrid systems in which trapped ions sympathetically cool and interact with nanomechanical oscillators or electric circuits.

\begin{figure}[!tb]
\includegraphics[width=0.83\columnwidth]{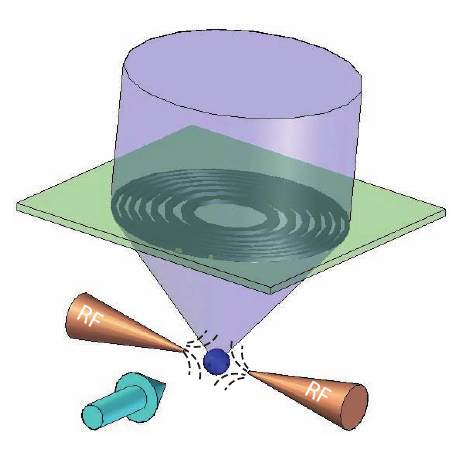}
\caption{Schematic of the experimental apparatus. A single Yb$^+$ ion is confined at the node of an RF electric quadrupole (dashed lines) formed between two needle electrodes. The ion is laser cooled using a $\lambda=369.5$ nm laser beam (arrow) and the resulting fluorescence is collimated using an in-vacuum phase Fresnel lens. This light is subsequently imaged onto a cooled CCD camera (not shown).}
\label{schematic}
\end{figure}

Existing trapped ion thermometry techniques in the millikelvin regime are spectroscopic and rely on varying the cooling laser frequency to obtain a measurement of Doppler broadening due to ion motion\cite{Wineland-Itano-79, Wineland-Itano-81}. This approach pushes the system out of thermal equilibrium, with the variation in laser frequency affecting the cooling rate and subsequently changing the temperature over the course of the measurement. Recoil heating of the ion increases its temperature and reduces the peak fluorescence when the laser detuning is less than half the linewidth from resonance. At laser detunings larger than half the linewidth the cooling rate is reduced causing an increase in the temperature and the associated Doppler linewidth. Both of these effects produce a systematic error that leads to an overestimation of the temperature. Time resolved techniques to quantify the effects of non-equilibrium laser cooling dynamics on the fluorescence signal have been demonstrated\cite{Epstein-07,Daniilidis-11} but depend on numerous simplifying assumptions or extensive simulation\cite{Wesenberg-07}. The Doppler approach is also limited to measurement of the ion motion in a single dimension fixed along the axis of the laser beam. The spectroscopic linewidth of the transition limits the minimum detectable temperature by the Doppler technique. These deficiencies are less of an issue in traps with low anomalous heating rates\cite{Labaziewicz-08} where sub-millikelvin resolved sideband thermometry is feasible.

We demonstrate an alternative approach to trapped ion thermometry which probes the spatial distribution rather than the velocity distribution. A key enabling technology of this technique is our recent demonstration of high resolution imaging of trapped ions with spot sizes as small as 373 nm $1/e^2$ radius (440 nm FWHM) \cite{Jechow-Kielpinski-2011}. Such high resolution enables us to measure temperatures of a few millikelvin by spatial thermometry and investigate phenomena inaccessible through spectroscopic thermometry. For a particle in a harmonic potential the spatial extent is dependent on the temperature and the trapping frequency. Imaging an ion thus provides a measure of its temperature in two dimensions and under steady-state conditions. Other trapped ion systems typically have imaging resolutions of several $\mu$m and are thus limited to a spatial temperature sensitivity of a few Kelvin. Our accuracy of $\pm$5mK is limited by uncertainty in the imaging resolution while the precision of $\pm$1mK is limited by image signal to noise.

The apparatus is similar to that described previously\cite{Jechow-Kielpinski-2011}. $^{174}$Yb$^+$ ions were generated by isotope-selective photoionisation of a Yb beam and were trapped in an RF electric quadrupole field formed by applying a potential $V_0\cos(\Omega_{RF}t)$, $\Omega_{RF}/2\pi =20$ MHz between two needles separated by 300 $\mu$m. We obtained ion motional frequencies between 500 kHz and 1500 kHz. Residual electric fields at the trap center were canceled by applying DC voltages to compensation electrodes. The ions were laser cooled using $\lambda=369.5$ nm light resonant with the S$_{1/2}$ to P$_{1/2}$ transition. The frequency of the cooling laser was stabilized using dichroic atomic vapor laser locking (DAVLL) \cite{Streed-Kielpinski-2008}. Detuning the laser frequency changed the DAVLL offset which in turn could be calibrated using a Fabry-Perot interferometer with a known free spectral range. The ion fluorescence was collected using a phase Fresnel lens (PFL) mounted in the ultra-high vacuum chamber, close to the trapping region \cite{Streed-Kielpinski-2011}. The PFL has an numerical aperture (NA) of 0.64 which covered 12$\%$ of the solid angle and collected 4.5$\%$ of the total emission. The collimated fluorescence was imaged with a magnification of $596\pm22$ on to a cooled CCD camera.

\begin{figure}[!tb]
\includegraphics[width=0.26\columnwidth]{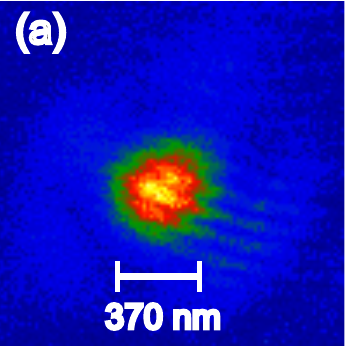}
\includegraphics[width= 0.26\columnwidth]{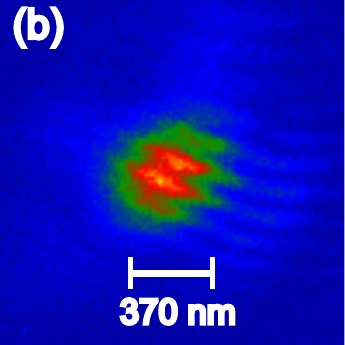}
\includegraphics[width= 0.26\columnwidth]{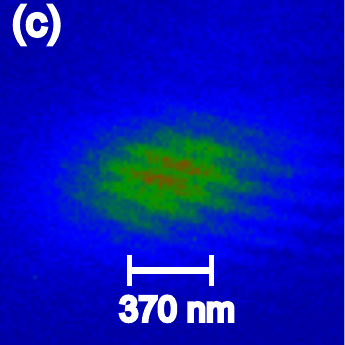}
\caption{Images of a trapped ion for three different external heating rates. Temperatures were calculated using equation \ref{eq:1} from the horizontal (laser cooling direction) ion spot radii. In the case of no (a), low (b) and high (c) heating rates we calculate temperatures of $5\pm5$ mK, $30\pm9$ mK and $162\pm14$ mK. Errors were dominated by systematic uncertainty.}
\label{ion image}
\end{figure}

We initially made a qualitative investigation of the spatial thermometry technique by controllably heating the ion. To heat the ion, we applied voltage with a white noise spectrum and a bandwidth encompassing all motional frequencies of the ion to one of the compensation needles. This allowed us to investigate different temperature ranges in a controlled fashion. Figure \ref{ion image} shows ion images for three different levels of heating; (a) no external heating, (b) low external heating and (c) high external heating (3.3x the noise voltage of b). There is a clear increase in the ion spot size with the increase in ion temperature.

From these high resolution images, we can quantitatively calculate the temperature of the ion. We fit a two dimensional Gaussian to the ion image in order to quantify the ion image ${1}/{e^2}$ radii $w_{x,y}$ for horizontal (x) and vertical (y) axes. The vertical axis of the image is parallel to one of the trap axes. The motional frequencies along the other two trap axes are degenerate. Assuming a thermal distribution \cite{Fundamentals-Of-Physics}, the ion temperature along the x axis is given by
\begin{equation}
T_x=\frac{m\pi^2{\nu_x}^2 \langle x^2 \rangle}{k_B}
\label{eq:1}
\end{equation}
where $m$ is the mass and $k_B$ is Boltzmann's constant and $\langle x^2 \rangle$ is the variance of the ion spatial distribution along the x axis. Owing to the degeneracy of the other two trap axes, a similar equation holds for y.

In the low temperature limit, the finite resolution of the imaging system dominates the observed ion image radius. Using our knowledge of the imaging system, we can estimate the variance of the ion spatial distribution and hence the temperature. To extract the variance of the ion spatial distribution we deconvolved the measured ion image radius with the imaging resolution ($w_i$) according to $\langle x^2 \rangle=\frac{{w_x}^2-{w_i}^2}{4}$. Uncertainty in the imaging resolution leads to a systematic uncertainty in the variance of the ion spatial distribution and thus the measured temperature. We evaluate the uncertainty by considering upper and lower bounds on our imaging resolution. An ideal imaging system has a minimum $1/e^2$ radius resolution of $0.43 \lambda/NA$. In our system this gives a diffraction limited lower bound on the imaging resolution of $249$ nm. The smallest spot radius observed in the horizontal axis was $373$ nm, giving an experimental resolution limited upper bound. Deconvolving the radii of the ion image shown in Fig \ref{ion image} a) with the lower resolution bound gives a temperature of $8\pm1$ mK. Here the statistical error of 1 mK arises from the error in estimation of $w_x$ and sets the resolution limit of spatial thermometry. The spot size of this image is the smallest we have observed, so deconvolution with the upper bound of imaging resolution gives a temperature estimate of approximately zero. The deconvolution procedure significantly improved the accuracy of our technique in the few mK temperature range. This systematic uncertainty is the primary limit to the accuracy of spatial thermometry. In the following, we state systematic uncertainties for imaging temperatures unless indicated otherwise.

We compared the imaging thermometry technique to the established Doppler spectroscopic thermometry technique. The ion fluorescence spectrum under controlled heating was measured by sweeping the laser frequency $\nu_L$ over the cooling transition for various levels of applied noise voltage. When the laser was detuned to frequencies above resonance, the ion was rapidly heated causing a sharp decrease in the fluorescence. The half Voigt profile that is conventionally used to fit the spectral response \cite{Kielpinski-Kartner-2006} was not able to discriminate between Lorentzian and Gaussian linewidth contributions. Instead we fit the data to a Lorentzian with a smoothed step function cutoff
\begin{equation}
\frac{1}{{\delta\nu_L}^2 + \left(\Gamma_T/2\right)^2}\cdot \arctan\left(\frac{\delta\nu_L}{\Delta_L}\right),
\label{eq:3}
\end{equation}
where $\delta\nu_L$  is the laser frequency detuning, $\Delta_L$ is the linewidth of the cooling laser and $\Gamma_T$ is the FWHM spectral linewidth of the ion.
To determine the temperature using the spectroscopic thermometry technique, the total spectral linewidth ($\Gamma_T$) was deconvolved into its two constituent components \cite{Olivero-Longbothum-1977}. The Gaussian component ($\Gamma_G$) is related to the temperature of the ion by $T=\frac{m}{2k_B}\left(\Gamma_G\lambda\right)^2$ \cite{Laser Spectroscopy} and the Lorentzian component ($\Gamma_L$) is related to the linewidth of the ion at zero temperature.

A systematic uncertainty in the temperature estimated by the spectroscopic thermometry technique arises from unresolved contributions to the Lorentzian component of the spectroscopic linewidth. We calculate the upper bound for the spectroscopic temperature by assuming that $\Gamma_L$ is the natural linewidth, resulting in a maximum contribution from $\Gamma_G$. This bound assumes that broadening mechanisms such as Zeeman splitting, saturation and micromotion have no contribution to $\Gamma_L$. The lower bound on the spectroscopic temperature is calculated by assuming $\Gamma_L$ is equal to the smallest experimentally observed linewidth.

\begin{figure}[tc]
\includegraphics[width=0.83\columnwidth]{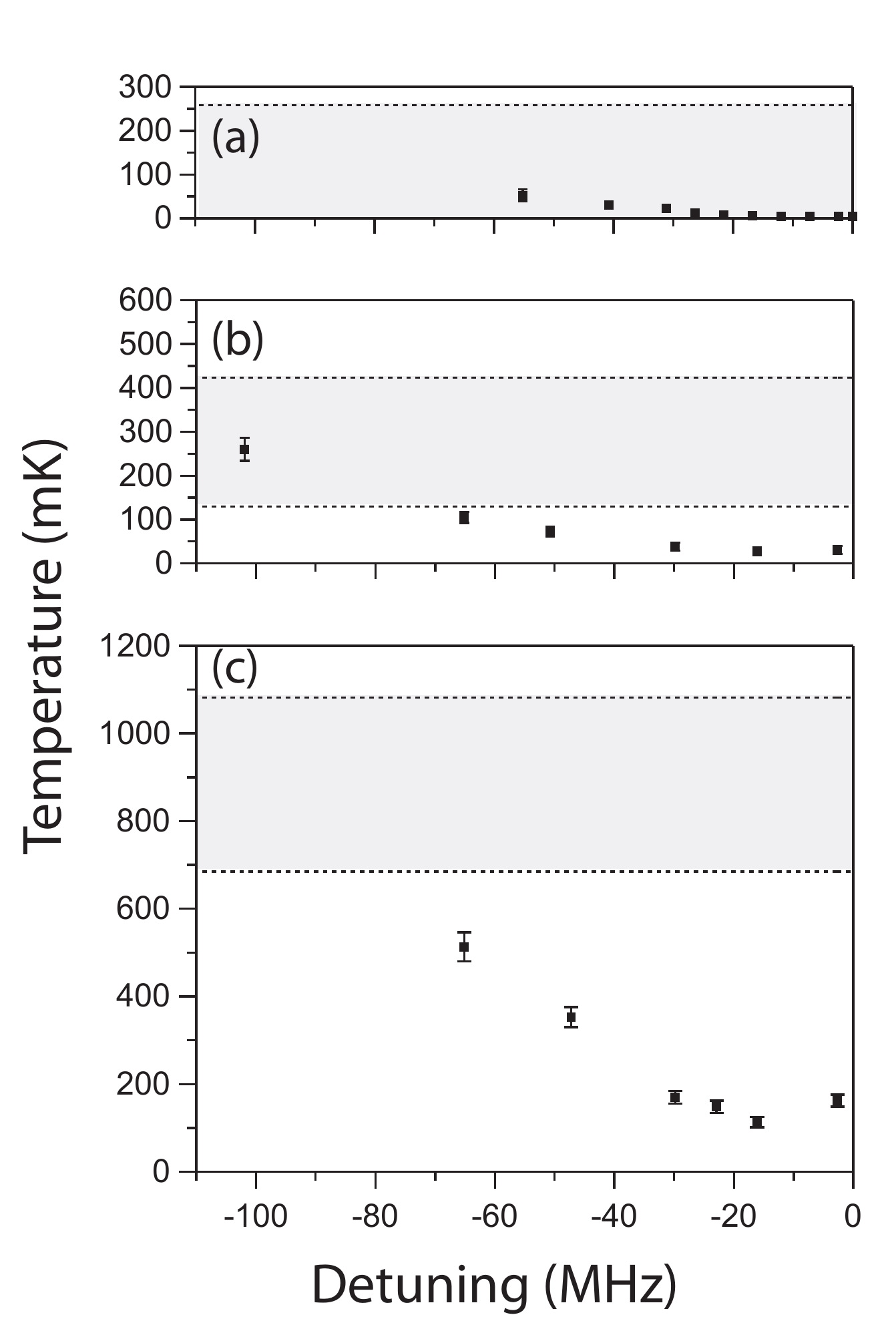}
\caption{Ion temperature dependence on cooling laser detuning for three different external heating rates. Both spectroscopic and imaging temperatures for (a) no, (b) low, and (c) high external heating are shown. The uncertainty in the imaging temperature is dominated by the systematic uncertainty in the correction for ion spatial extent by the imaging resolution. The spectroscopic temperature 1$\sigma$ uncertainty band for each of the three external heating rates is shown as the gray band with the upper and lower bound depicted as a dashed line.}
\label{temp graph}
\end{figure}

Figure \ref{temp graph} compares the spectroscopic temperature results to those from the imaging method. The same three external heating rates as in Fig \ref{ion image} are used and imaging temperatures are obtained for several cooling laser detunings. The imaging temperatures at each detuning were calculated from the fitted ion spot sizes in the horizontal (cooling laser) direction. In all cases the imaging thermometry is far more precise and appears to be more accurate. The temperatures calculated by spectroscopic thermometry show large systematic uncertainties in all cases. The spectroscopic thermometry technique overestimates the temperature as expected from considerations of laser cooling dynamics.

Spatial thermometry shows dynamics that lead to systematic error in spectral thermometry. At large detunings an increase in temperature was observed as the scattering rate dropped, reducing the efficiency of Doppler cooling. When the cooling laser was close to resonance an increase in the temperature occurred due to increased recoil heating. In Fig \ref{temp graph} a), with no applied heating, a minimum in the ion imaging temperature was observed at $\sim15$ MHz corresponding to the expected optimum detuning of $\Gamma_T/2$. The systematic uncertainty in the imaging temperature dominates the measurement accuracies for temperatures close to the Doppler cooling limit. As external heating is applied and the temperature of the ion increases, the systematic uncertainty becomes negligible. In Figure \ref{temp graph} (c), imaging temperatures for detunings greater than 70 MHz were not recoverable due to the configuration of the imaging system.

\begin{figure}[!hc]
\includegraphics[width=0.83\columnwidth]{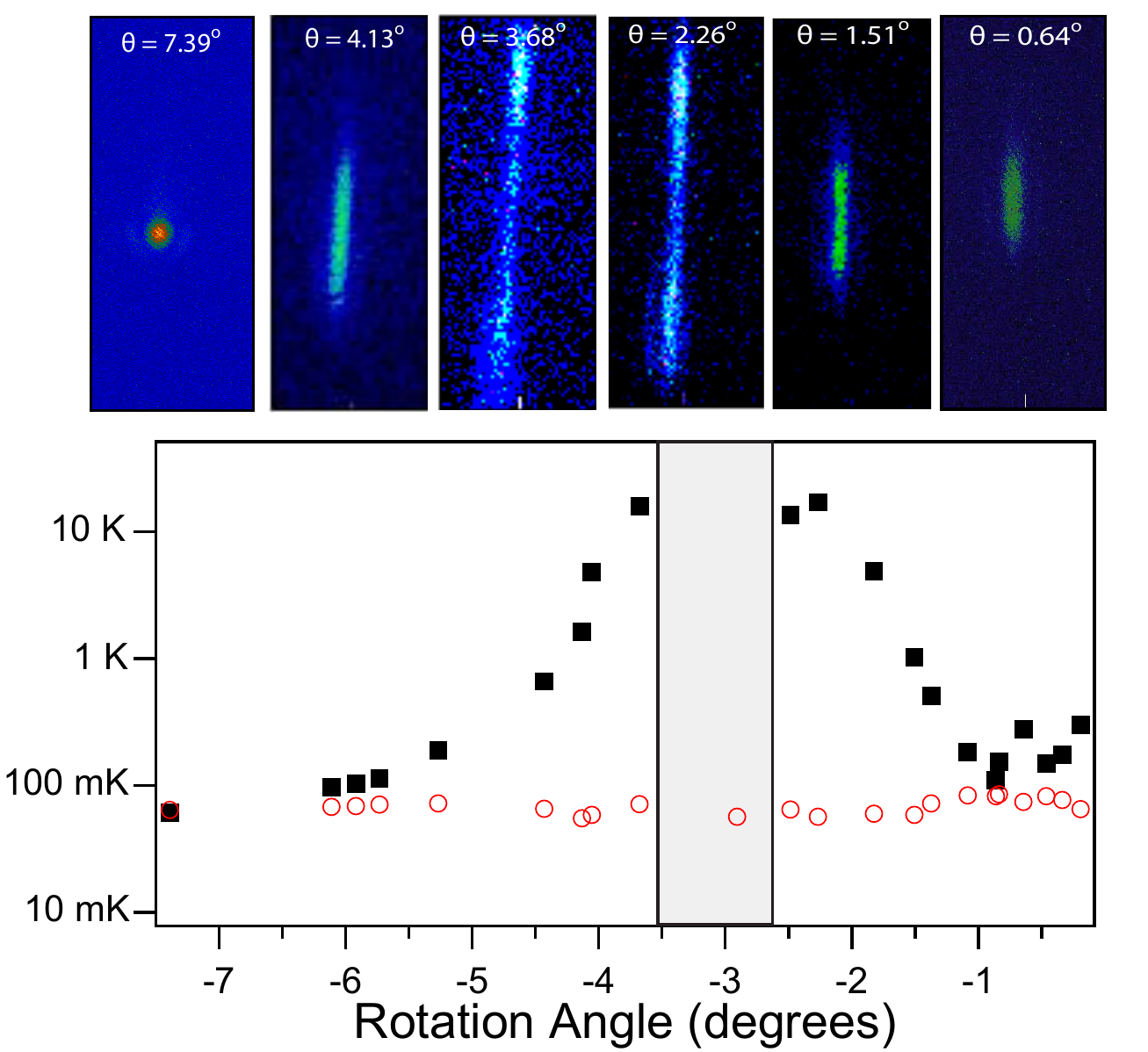}
\caption{Spatial thermometry of a single ion under highly anisotropic laser cooling dynamics. Top: ion images for several rotation angles. The contrast of the ion images has been adjusted for ease of viewing. Bottom: temperature along weak-cooling axis (filled black squares) and strong-cooling axis (open red circles). For rotation angles between -3.5$^{\circ}$ and -2.5$^{\circ}$ (gray shaded area) the size of the ion image is larger than the camera chip along the vertical direction. A temperature difference of a factor of 1000 was sustained without any adverse effects on the trapping. Error bars are smaller than the data points.}
\label{rotation vs dc bias}
\end{figure}

The spatial thermometry technique allows measurement of temperature along two dimensions even when the ion dynamics are highly anisotropic. Figure \ref{rotation vs dc bias} shows the effect of highly anisotropic laser cooling dynamics for a single ion. The cooling laser was controllably decoupled from one axis of motion by applying a DC bias to the RF needles, which caused rotation of the trap axes. When the cooling laser is decoupled from one axis of motion, the cooling power is reduced while there is no reduction in recoil heating along this axis. This rotation is believed to occur due to the difference in the RF and DC grounding.  Figure \ref{rotation vs dc bias} shows the temperature of the ion as a function of trap axis rotation demonstrating a large disparity between the horizontal and vertical ion temperatures. Ion images in Fig \ref{rotation vs dc bias} show the ion spatial extent at selected rotations.  A temperature difference of a factor of 1000 was observed without any adverse effects on the trapping. This temperature difference was achieved with a maximum trap aspect ratio of 2.25. The angular width of the feature in Fig \ref{rotation vs dc bias} b) was much larger than the angular width of the beam used to cool the ions ($0.3^{\circ}$). Spectral thermometry along the vertical axis would require the laser direction to have a projection along that axis, inducing laser cooling. Hence, the anisotropic dynamics studied here can only be observed by spatial thermometry.

High-resolution imaging of trapped ions was used to realize steady state millikelvin thermometry and investigate phenomena inaccessible through conventional Doppler spectroscopy. We achieved an accuracy of $\pm5$ mK with temperature resolutions of $\pm1$ mK, limited by the imaging system resolution. This spatial thermometry technique is independent of laser cooling dynamics and has been used to observe the dependence of ion temperature with laser detuning for three different heating rates. The variation of temperature with cooling laser detuning is a prominent source of systematic error in Doppler thermometry and plays an important role in estimating heating rates \cite{Wesenberg-07}. By rotating the trap axes with respect to the cooling laser we investigated the ion dynamics under highly anisotropic laser cooling. Temperatures of $<$ 60mK and $>$15K were simultaneously measured along two motional axes of a single trapped ion. The latter temperature is easily accessible by optical cryostats and demonstrates the utility of this technique for investigations of sympathetic cooling in proposed hybrid trapped ion systems with nanomechanical oscillators\cite{Tian-2004,Hensinger-2005} or resonant electrical circuits\cite{Rabl-2010}. Investigations of highly localized variations in anomalous heating rates observed in micro-fabricated ion traps developed for QIP\cite{Daniilidis-11} suggests a strongly anisotropic process that could be better resolved using the multi-dimensional aspects of spatial thermometry.

This work is funded by the Australian Research Council under DP0773354 (DK), DP0877936 (ES), and FF0458313 (H. Wiseman, Federation Fellowship), as well as the US Air Force Office of Scientific Research (FA2386-09-1-4015). AJ is supported by a Griffith University Postdoctoral Fellowship.

\end{document}